\documentstyle[sprocl,epsfig]{article}

\def\be{\begin{equation}}
\def\ee{\end{equation}}
\def\bea{\begin{eqnarray}}
\def\eea{\end{eqnarray}}
\newcommand{\alfa}{\mbox{$\alpha$}}

\newcommand{\aabha}{\mbox{\tt SAMBHA}}

\newcommand{\QQ}{\mbox{$q^2$}}
\newcommand{\Da}{\mbox{$\Delta\alfa$}}
\newcommand{\mzz}{\mbox{$m_Z^2$}}

\newcommand{\dd}{{\mathrm d}}

\begin{document}

\title{MEASURING THE RUNNING OF THE ELECTROMAGNETIC COUPLING ALPHA IN SMALL ANGLE BHABHA SCATTERING~\footnote{Work done in collaboration with A.Arbuzov, D.Haidt, C. Matteuzzi and M. Paganoni}}

\author{ LUCA TRENTADUE \footnote{On leave of absence from Dipartimento di Fisica, Universit\'a di Parma, 43100, Parma, Italy}}
\address{Department of Physics, CERN Theory Division, 1211 Geneva 23, Switzerland\\and\\INFN, Gruppo Collegato di Parma,  I-43100 Parma, Italy}


\maketitle\abstracts{We propose a method to determine the running of $\alpha_{QED}$
from the measurement of small-angle Bhabha scattering. The method is suited to high statistics experiments at $e^{+} e^{-}$ colliders, which are equipped with luminometers in the appropriate angular region. We present a new simulation code predicting small-angle Bhabha scattering. A detailed description of this idea can be found in ref.[1]}

\section{Introduction}
The electroweak Standard Model $SU(2)\otimes U(1)$ contains Quantum Electrodynamics (QED) as a constitutive 
part. The running of the electromagnetic
coupling \alfa\ is determined by the theory as
\begin{equation}
  \alfa(\QQ) = \frac{\alfa(0)}{1-\Da(\QQ)}, \label{eq:Da}
\end{equation}
where \alfa(0)\,=\,$\alpha_0$ is the Sommerfeld
fine structure constant, which has been measured to 
a precision of 3.7$\times$10$^{-9}$\ \cite{pdg00}; $\Delta\alpha(q^2)$
positive arises from loop contributions to
the photon propagator. The numerical prediction of electro\-weak observables
involves the know\-ledge of \alfa(\QQ), usually  for \QQ \,$\neq$\, 0. For 
instance, the know\-ledge of \alfa(\mzz) is relevant 
to the evaluation of quantities 
measured by the LEP experiments. This is achieved by evolving \alfa\ from 
\QQ\,=\,0\ up to the $Z$-mass scale \QQ\, =\, \mzz. 
The evolution expressed by the quantity \Da\ receives contributions from
leptons, hadrons and the gauge bosons. The hadronic contribution to the
vacuum polarization, which 
cannot be calculated from first principles, is estimated with the help of a 
dispersion integral and evaluated \cite{fj} by using total cross section 
measurements of $e^+e^-\rightarrow$ hadrons at low energies. Therefore, any evolved 
value \alfa(\QQ)\, particularly  for $|q^2|>4m_{\pi}^2$, 
is affected by uncertainties 
originating from hadronic contributions.  
The uncertainty on $\alfa(\mzz)^{-1}$ induced by these 
data is as small as $\pm$0.09 \cite{fj}; nevertheless it turned out 
\cite{hhm} that this
limits the accurate prediction of electroweak quantities within the 
Standard Model, parti\-cularly for the prediction of the Higgs mass. 

While waiting for improved measurements from BEPC, VEPP-4M and DAFNE as input
to the dispersion integral, intense efforts are
made to improve on estimating the hadronic shift $\Da_{had}$,
as for instance \cite{fj99}-\cite{pp}, and to find alternative ways of measuring
\alfa\ itself. Attempts have been made to measure \alfa(\QQ) directly, using 
data at various energies, such as measuring the ratio
of $e^+e^-\gamma$/$e^+e^-$ \cite{topaz} or more directly the angular 
distribution of Bhabha scattering \cite{L3}.
In this article the running of \alfa\ is studied using small-angle Bhabha 
scattering. This process provides unique information on the QED coupling 
constant \alfa\ at low {\it space-like} momentum transfer $t=-|\QQ|$, where
\begin{equation}
 t = -\frac{1}{2}\ s\ (1-\cos\theta) \label{eq:t}
\end{equation}
is related to the total invariant energy $\sqrt{s}$ and to the scattering 
angle $\theta$ of the final-state electron. The small-angle region 
has the virtue of giving access to values of \alfa($t$)\ without being affected
by weak contributions. The cross section can be theoretically calculated 
with a precision at the per mille level. It is dominated by the photonic
$t$ channel exchange and the non-QED contributions have been
computed~\cite{luca} and are of the order of 10$^{-4}$;
in particular, contributions from boxes with two weak bosons are
safely negligible.

In general, the Bhabha cross section is computed from the entire set of 
gauge-invariant amplitudes in both the $s$ and $t$ channels. Consequently, two invariant 
scales $s$ and $t$ govern the process. The different amplitudes are functions of both
$s$ and $t$ and also the QED coupling $\alpha$ appears as $\alpha(s)$ resp. $\alpha(t)$
\cite{brodsky}. However, the restriction of Bhabha scattering to the kinematic regime of 
small angles results in a considerable simplification, since 
the $s$ channel then gives only a negligible contribution, as is quantitatively 
demonstrated~\cite{lavoro}. Thus, the measurement of the angular distribution 
allows us indeed to verify directly the running of the coupling $\alpha(t)$. 
For the actual calculations, $\theta\gg m_e/E_{beam}$ and 
$E_{beam}\gg m_e$ must be satisfied.
Obviously, in order to mani\-fest the running, the experimental precision must be ad\-equate. 
This idea can be realized by high-statistics experiments at $e^+e^-$ colliders
equipped with finely segmented luminometers, in particular by 
the LEP experiments, given their large event samples, by SLC and future Li\-near 
Colliders. The relevant luminometers
cover the $t$-range from a few GeV$^2$ to order 100 GeV$^2$. 
The analysis 
follows closely the procedure adopted in  the luminosity measurement, which 
is described in detail~\cite{LEP_report}, and
elaborates on the additional aspect related to the measurement of a 
differential quantity. To this
aim the luminosity detector must have a sufficiently large angular acceptance 
and adequate fine segmentation. The variable $t$ (eq.~\ref{eq:t}) is 
reconstructed on an event-by-event basis. 
The cross section for the process $e^+e^- \rightarrow e^+e^-$ can be conveniently decomposed 
into three factors~:
\begin{equation}
\frac{\dd \sigma}{\dd t} = \frac{\dd \sigma^0}{\dd t}
                     \left(\frac{\alpha(t)}{\alpha(0)}\right)^2
                     (1+\Delta r(t)) \label{eq:meth}.
\end{equation}
All three factors are predicted to a precision of 0.1\% or better. 
The first factor on the right-hand side refers to the effective Bhabha Born cross section, 
including soft and virtual photons~\cite{luca},
which is precisely known, and accounts for the strongest dependence on $t$. 
The vacuum-polarization effect in the leading photon $t$ channel exchange
is incorporated in the running of $\alpha$ and gives rise to the squared factor in
eq.~\ref{eq:meth}.
The third factor, $\Delta r(t)$, 
collects all the remaining real (in particular collinear) and virtual radiative effects
not incorporated in the running of $\alpha$. 
The experimental data after correction for detector effects have to be compared
with eq.~\ref{eq:meth}. The $t$ dependence is rather steep, 
thus migration effects may need attention.

This goal is achieved by using a newly developed program based on the already existing
semianalytical code {\tt NLLBHA} \cite{luca,nllbha}. A complete bibliography of this
code called \aabha\ is given in ref.[1].  
It is convenient to confront the fully corrected measured cross section
with the Bhabha cross section, including radiative corrections in the factorized 
form given by eq.~\ref{eq:meth}. 
The physical cross section is infrared safe \cite{luca}.
This decomposition is neither unique nor dictated by a compelling physical reason;
rather it allows the separation of the different sources of $t$ dependence in
a transparent way without introducing any additional theoretical uncertainty.
The va\-rious factors as well as the relative accuracy of the QED corrections ( see for example refs [15]-[16] )
and the MonteCarlo tool to implement them are discussed in ref.[1]. 
\section{Conclusions}
A novel approach to access directly and to measure the running of 
$\alpha$ in the space-like region
is proposed. It consists in analy\-sing small-angle Bhabha scattering.
Depending on the particular angular detector coverage and on the energy of the
beams, it allows a sizeable range of the $t$ variable to be covered.
The feasibility of the method has been put in evidence by the use of a new
tool, \aabha\ , to calculate the small-angle Bhabha differential cross section with 
a theoretical accuracy of better than 0.1\%.
The information obtained in the $t$ channel can be compared with the existing 
results of the $s$ channel measurements. This represents a complementary approach,
which is direct, transparent and based only on QED interactions and furthermore
free of some of the drawbacks inherent in the $s$ channel methods. 
The method outlined can be readily applied to the experiments at LEP and SLC.
It can also be exploited by future $e^+e^-$ colliders as well as by existing lower
energy machines. As an example the $q^2$ values $[GeV^2]$
as the $E_{Center\; of\; Mass }$ energy  varies between 0 and $10^3\;GeV$ within a range of acceptance for the emitted radiation between the angles from 1.8 to 7.2 degrees range between 0 and $4\cdot 10^3 \;GeV^2$.

An extremely precise measurement of the loop contributions to the QED running coupling $\Delta\alpha(t)$ 
for small values of $t$ may be envisaged with a dedicated luminometer 
even at low machine energies.

\end{document}